\documentclass[useAMS,usenatbib]{mn2e}
\usepackage{amsmath,amstext,amsgen,amsbsy,amsopn,amsfonts,theorem}

\usepackage{graphicx}
\usepackage{xspace}
\usepackage{amsmath}
\usepackage{hyperref}
\usepackage{framed}
\usepackage{txfonts}
\usepackage{epstopdf}

\makeatletter
\def\mr@ignsp#1 {\ifx\:#1\@empty\else #1\expandafter\mr@ignsp\fi}%
\newcommand{\multiref}[1]{\begingroup
\xdef\mr@no@sparg{\expandafter\mr@ignsp#1 \: }%
\def\mr@comma{}%
\@for\mr@refs:=\mr@no@sparg\do{\mr@comma\def\mr@comma{,}\ref{\mr@refs}}%
\endgroup}
\makeatother

\newcommand{\hypref}[2]{\ifx\href\asklfhas #2\else\href{#1}{#2}\fi}
\newcommand{\Secref}[1]{Section~\multiref{#1}}
\newcommand{\secref}[1]{Sec.~\multiref{#1}}

\newcommand{\Figref}[1]{Figure~\multiref{#1}}
\newcommand{\figref}[1]{Fig.~\multiref{#1}}
\newcommand{\Eqref}[1]{Equation~(\multiref{#1})}
\renewcommand{\eqref}[1]{(\multiref{#1})}

\def\[{\begin{equation}}
\def\]{\end{equation}}
\def\<{\begin{eqnarray}}
\def\>{\end{eqnarray}}

\newcommand{\HI}{{\rm HI}}

\newcommand{\HeI}{{\rm HeI}}
\newcommand{\HeII}{{\rm HeII}}
\newcommand{\nh}{{n_{\rm H}}}
\newcommand{\nhi}{{n_\HI}}
\newcommand{\NHI}{N_{\HI}}
\newcommand{\NH}{N_{\rm H}}
\newcommand{\pcms}{{\rm cm}^{-2}}
\newcommand{\Msun}{{\rm M}_\odot}
\newcommand{\der}{{\rm d}}

\title[The Physics and Environment of Lyman Limit Systems]{Investigating the Physics and Environment of Lyman Limit Systems in Cosmological Simulations}

\author[D. Erkal]
  {Denis Erkal$^1,^2$\thanks{derkal@ast.cam.ac.uk} \\
  $^1$Institute of Astronomy, Madingley Road, Cambridge, CB3 0HA, UK\\
  $^2$Department of Physics, The University of Chicago, Chicago, IL 60637,
USA}

\begin{document}

\label{firstpage}

\maketitle

\begin{abstract}
In this work, I investigate the properties of Lyman limit systems (LLSs) using
state-of-the-art zoom-in cosmological galaxy formation simulations with on the fly
radiative transfer, which includes both the cosmic UV background (UVB) and local stellar
sources. I compare the simulation results to observations of the incidence frequency of
LLSs and the HI column density distribution function over the redshift range $z=2-5$ and
find good agreement. I explore the connection between LLSs and their host halos and find
that LLSs reside in halos with a wide range of halo masses with a nearly constant
covering fraction within a virial radius. Over the range $z=2-5$, I find that more than
half of the LLSs reside in halos with $M < 10^{10}h^{-1}M_\odot$, indicating that
absorption line studies of LLSs can probe these low-mass galaxies which H$_2$-based star
formation models predict to have very little star formation. I study the physical state
of individual LLSs and test a simple model (Schaye 2001) which encapsulates many of their
properties. I confirm that LLSs have a characteristic absorption length given by the
Jeans length and that they are in photoionization equilibrium at low column densities.
Finally, I investigate the self-shielding of LLSs to the UVB and explore how the
non-sphericity of LLSs affects the photoionization rate at a given $N_{\rm HI}$. I find
that at $z\approx 3$, LLSs have an optical depth of unity at a column density of $\sim
10^{18} {\rm cm}^{-2}$ and that this is the column density which characterizes the onset
of self-shielding.
\end{abstract}


\section{Introduction}

Lyman limit systems are a special class of Ly$\alpha$ absorbers which span a range of
column densities: $\NHI = 1.6\times10^{17}\pcms - 2\times10^{20}\pcms$. The lower limit
is defined by the column density which gives an optical depth of unity at the Lyman limit
and the upper limit is defined by the transition to Damped Ly$\alpha$ (DLA) systems which
are mostly neutral. They are primarily observed through quasar absorption lines although
their absorption features have also been seen in the spectra of gamma-ray bursts. See
\cite{rauch_1998,meiksin_2009} for reviews of Ly$\alpha$ absorbers and
\cite{wolfe_et_al_2005} for a review of DLAs.

Observations of LLSs and DLAs in the high-redshift universe provide a fertile ground for
comparison with theoretical work. They give a unique window into the high-redshift
universe since the quasar absorption line observations provide an area-weighted survey of
these absorbers across a large range of redshifts which makes them especially simple to
compare with simulations.

While the absorption line studies provide rich statistics of these systems when averaged
over many lines of sight, it is difficult to deduce the environment in which individual
absorbers reside. The main goal of this work is to understand the environment of LLSs, as
well the physical mechanisms which control their properties. Many groups have studied the
properties of LLSs in simulations of varying mass resolution and with many of the
physical mechanisms which affect LLSs
\citep{kohler_gnedin_2007,altay_et_al_2011,mcquinn_et_al_2011,fumagalli_et_al_2011,yajima_et_al_2011,rahmati_et_al_2012,rahmati_et_al_2013,rahmati_et_al_2013_b}.
The simulations in this work have a relatively high mass resolution of $1.5\times 10^5
{\rm h}^{-1} M_\odot$, allowing us to study lower mass halos, $M < 10^9 {\rm h}^{-1}
M_\odot$, than has previously been achieved. This mass range is especially interesting
since H$_2$-based star formation models indicate that these halos will not form stars
\citep{gnedin_kravtsov_2010,kuhlen_et_al_2013} and hence they may only be detectable
using absorption line studies.

In addition to studying the halos in which LLSs reside, I will use these simulations to
study the self-shielding of LLSs to the UVB. LLSs are defined as having an optical depth
greater than unity to radiation at the Lyman limit, i.e. $\NHI = 10^{17.2}\pcms$. The
column density at which this self-shielding becomes effective is important since it
controls the turnover of the HI column density distribution as was shown in
\cite{altay_et_al_2011,mcquinn_et_al_2011,rahmati_et_al_2012}. In
\Secref{sec:effective_shielding}, I will show that due to the physical properties and
anisotropic shielding of LLSs, as well as the spectrum of the UVB, a column density of
$\NHI \sim 10^{18}\pcms$ is needed to shield against the UVB with an optical depth of
unity at $z\approx3$.

This paper is arranged as follows. In \Secref{sec:simulation_description}, I discuss the
simulations used in this paper. Next, I compare the simulation results to quasar
absorption line observations of the high-redshift universe in \Secref{sec:cdd} and find
that the simulations qualitatively reproduce the features seen in observations. In
\Secref{sec:LLS_and_Halos}, I explore the relation between LLSs and their host halos and
find that LLSs reside in halos with a large range of masses but that there is a cutoff at
low mass which is similar to the cutoff due to photoheating from the UVB. In
\Secref{sec:individual_LLS}, I investigate the physical mechanisms of individual LLSs and
test a simple model for LLSs developed in \cite{schaye_2001}. In \Secref{sec:anisotropy},
I study the anisotropy of LLSs and how this affects their self-shielding properties. In
\Secref{sec:effective_shielding}, I discuss how the physical properties of LLSs and the
spectral shape of the UVB affect the amount of self-shielding in these systems. In
\Secref{sec:other_works}, I compare the results from this work to some recent works on
LLSs. Finally, I conclude in \Secref{sec:conclusion}.

\section{Simulations} \label{sec:simulation_description}

In this work, I have used the simulation described in \cite{zemp_et_al_2012}, carried out
using the Adaptive Refinement Tree (ART) code
\citep{kravtsov_1999,kravtsov_et_al_2002,rudd_et_al_2008}. The code has adaptive mesh
refinement which gives a large dynamic range in spatial scale. These simulations follow
five different Lagrangian regions, each of five virial radii around a system which
evolves into a typical halo of an $L_*$ galaxy (${\rm M} \approx 10^{12} \Msun$) at
$z=0$. These Lagrangian regions are embedded in a cube of size $25.6$ comoving $h^{-1}$
Mpc to model the tidal forces from surrounding structures. The outer region is coarsely
resolved with a uniform $256^3$ grid. The dark matter mass resolution is $1.5 \times
10^{5} h^{-1} \Msun$ in the high-resolution Lagrangian region and the baryonic mass
resolution varies from $\sim 10^3 \Msun$ to $\sim 10^6 \Msun$ depending on cell size and
density. The maximum spatial resolution is $195$ comoving $h^{-1}$pc. The cosmological
parameters used are similar to the WMAP7 parameters: $\Omega_M = 0.28$, $\Omega_B =
0.046$, $\sigma_8 = 0.82$, $h=0.7$, and $n_s = 0.96$.

These simulations include three-dimensional radiative transfer of UV radiation from the
UVB as well as from stars formed in the simulation. This is done with the Optically Thin
Variable Eddington Tensor (OTVET) approximation \citep{gnedin_abel_2001}. The
contribution from the UVB uses the model in \cite{haardt_madau_2001}, while the
contribution from local sources uses a Miller-Scalo IMF \citep{miller_scalo_1979} and the
shape of the spectrum from local sources comes from Starburst99 modeling
\cite{leitherer_et_al_1999} and is plotted in Figure 4 of \cite{ricotti_et_al_2002}. The
OTVET method in this work follows the transfer of radiation at 4 frequencies: at the
$\HI$, $\HeI$, and $\HeII$ ionization thresholds, as well as one to follow non-ionizing
radiation at 1000 \AA. The fidelity of this RT prescription was tested in
\cite{iliev_et_al_2006,iliev_et_al_2009} where it was found to work well except for some
numerical diffusion of ionization fronts. The prescription has subsequently been improved
and numerical diffusion has been almost completely eliminated \cite{gnedin_2014}. This
detailed and faithful radiative transfer allows us to model the self-shielding of LLSs
against the UVB. It is also important for understanding the effect of local sources on
LLSs since they arise in close proximity to galaxies.

These simulations include a self-consistent, non-equilibrium chemical network of hydrogen
and helium, including the effects of ionization from photoionization (corrected for
dust-shielding), collisional ionization, and radiative recombination
\cite{gnedin_kravtsov_2011}. The chemical network also self-consistently models $H_2$,
including the formation of molecular hydrogen in both primordial phase and on dust grains
\citep[see][for details]{gnedin_kravtsov_2011}. This physics includes the cooling and
physical mechanisms needed to correctly model the gas in LLSs and allows for a realistic
H$_2$-based star-formation model.

Finally, the simulations include thermal supernova feedback with an energy deposition of
$2\times10^{51}$ erg from Type Ia and Type II supernovae. This feedback prescription is
known to be inefficient since the supernova energy is deposited in cells with high
densities and relatively low temperatures which results in extremely efficient cooling.
While efficient feedback has been shown to increase the cross-section of LLSs
\citep[e.g.][]{faucher_et_al_2015,rahmati_et_al_2015} examining the effect of realistic
feedback is beyond the scope of this work. Note that since feedback also depends on the
mass of the host galaxy, the inclusion of more efficient feedback would also likely
affect the LLS cross-section versus halo mass which is explored below.

\section{Column Density Distribution and Incidence of LLSs} \label{sec:cdd}

Before delving into the properties of individual absorbers and their host halos, it is
useful to test how well the simulations are modeling the properties of LLSs by comparing
against observations. Two of the main statistics for LLSs measured by observers are the
number of LLSs per absorption length (the incidence frequency) and the number of systems
per unit absorption length per unit column density (the HI column density distribution).
The incidence frequency is written as,
\< l_{\rm LLS} = \frac{\der \mathcal{N}}{\der X}, \>
and the HI column density distribution is written as,
\< f(\NHI,z) = \frac{\der^2 \mathcal{N}}{\der\NHI \der X}, \>
where the absorption length is given by
\< \frac{\der X}{\der z} = \frac{H_0}{H(z)} (1+z)^2 .\>
These statistics are related since the HI column density distribution is the incidence
frequency per unit column density. The absorption length is defined this way so that
absorbers with a constant comoving number density and constant physical size have a
constant incidence frequency. Hence, any evolution in these quantities is due to
evolution in the cross-section of these systems, their number density, or a combination
of these two. Since LLSs reside in and around galaxies, their incidence can be written in
terms of the average LLS cross-section, $\sigma_{\rm LLS}(M,z)$, and the halo mass
function, $n(M,z)$, at redshift $z$ \citep[][]{gardner_et_al_1997}:
\< l_{\rm LLS} = \frac{c}{H_0}\int {\sigma}_{\rm LLS}(M,z) n(M,z) \der M
\label{eq:dNdX_halo_mf}.\>
Note that I will also consider the quantity $l_{\tau > \tau_0}$, which is the incidence
of systems with an optical depth greater than $\tau_0$ at the Lyman limit. Likewise, the
HI column density distribution can be written as
\< f(\NHI,z) = \frac{c}{H_0}\int \frac{\partial \sigma(\NHI,M,z)}{\partial \NHI} n(M,z)
\der M \label{eq:d2NdXdNHI_halo_mf} ,\>
where $\sigma(\NHI,M,z)$ is the average cross-section of absorbers with a column density
below $\NHI$ around halos of mass $M$.

\subsection{Observations of LLSs}

Observations of LLSs in the high-redshift universe are primarily made by using quasar
absorption lines. Since LLSs correspond to the flat portion of the curve of growth, their
column density is harder to determine than systems with lower or higher column densities.
The column densities of systems in the Ly$\alpha$ forest with $\NHI < 10^{17.2}\pcms$ can
be directly determined either from Voigt profile fits to the Ly$\alpha$ absorption, or
from fits to higher order Lyman transitions \citep[e.g.][]{rudie_et_al_2012}. For DLAs
and sub-DLAs, $\NHI > 10^{19} \pcms$, the natural line width of the Ly$\alpha$ transition
produces damping wings which make the column densities of these systems easy to determine
\citep[e.g.][]{wolfe_et_al_2005}. However, in the intermediate range, $10^{17.2} \pcms <
\NHI < 10^{19}\pcms$, the exact column density is difficult to measure and requires
precise observations of both the Ly$\alpha$ line and the Lyman limit break
\citep[e.g.][]{prochter_et_al_2010}. While the exact column density may be difficult to
determine in this range, the presence of an absorber with $\NHI > 10^{17.2}\pcms$ can be
inferred from the Lyman limit break. As a result, observers can more easily measure the
number of systems above a given threshold (typically $\NHI = 10^{17.2}\pcms$) which
provides an integral constraint on the HI column density distribution. In some works
\citep[i.e.][]{omeara_et_al_2012}, this counting is done for multiple thresholds which
can be used to constrain the column density distribution.

In \Figref{fig:dNdX}, I show observations of the incidence of LLSs over a variety of
redshifts. These come from \cite{prochaska_et_al_2010} and \cite{omeara_et_al_2012}. In
\Figref{fig:cdd}, I show the constraints on the HI column density distribution for LLSs
at $z\approx 2.4$ from \cite{omeara_et_al_2007} and \cite{omeara_et_al_2012}. Above $\NHI
= 10^{19}\pcms$ these constraints come from the detection of individual LLSs for which
the HI column density of each system can be determined. Between $\NHI = 10^{17.5} \pcms$
and $\NHI = 10^{19}\pcms$, the constraints are determined from $l_{\tau > 2}$. Below
$\NHI = 10^{17.5}\pcms$, the constraints are determined from the comparison of $l_{\tau >
2}$, $l_{\tau > 1}$, and $l_{\tau > 0.5}$ in \cite{omeara_et_al_2012}. See
\cite{omeara_et_al_2012} for a detailed discussion of these constraints.

\subsection{Measuring the Frequency and Column Density Distribution in Simulations}

Using a method similar to observations, the HI column density is computed by taking lines
of sight through the simulation, measuring the HI column density along these lines of
sight, and counting the number of absorbers in each column density bin. Observationally,
the HI column densities are determined by fitting profiles to the HI absorption lines. In
simulations, the HI column density can simply be integrated along lines of sight in the
three cartesian directions. Since systems in the simulation are randomly oriented with
respect to the simulation box, these lines of sight effectively probe random lines of
sight through systems in the simulation. This method gives the same HI column density as
fitting absorption lines as long as there are not multiple systems along each line of
sight.

In order to determine the column density at which these projection effects become
important, I considered lines of sight of various lengths along the cartesian directions.
These lines of sight were placed on a regular grid separated by 4 times the highest
resolution element, $781$ comoving $h^{-1}$ pc. This sampling fixes the number of lines
of sight taken through the simulation volume but does not affect the resolution along the
line of sight, which is controlled by the size of each cell. Along each line of sight, I
found the location of the cell with the maximum HI density and defined this to be the
center of the absorber. This definition will allow us to probe the environment which is
physically close to the absorber since we can take lines of sight originating from this
point. I then considered lines of sight of length 10kpc, 50kpc, 200kpc, and the full box
length, centered on the absorber. I found that while the 10kpc and 50kpc lines of sight
differed substantially below $\NHI = 10^{16}\pcms$, the 200kpc and full box lines of
sight showed fairly similar column densities (only 2.5\% of systems differed by more than
a factor of 2) indicating that the projection effects are not substantial for these
systems. In this work I will restrict the analysis to $\NHI > 10^{16.5}\pcms$ where
projection effects are even less important. This approach was also taken in
\cite{altay_et_al_2011} and \cite{rahmati_et_al_2012} where the projected column density
was used for systems with $\NHI > 10^{17}\pcms$ and $\NHI > 10^{16}\pcms$ respectively.
Note that these shorter lines of sight target gas associated with the absorber and will
also be used to measure quantities like the characteristic size of an absorber.

\subsection{LLS Incidence Frequency}

\begin{figure}
\centering
\includegraphics[width=8cm]{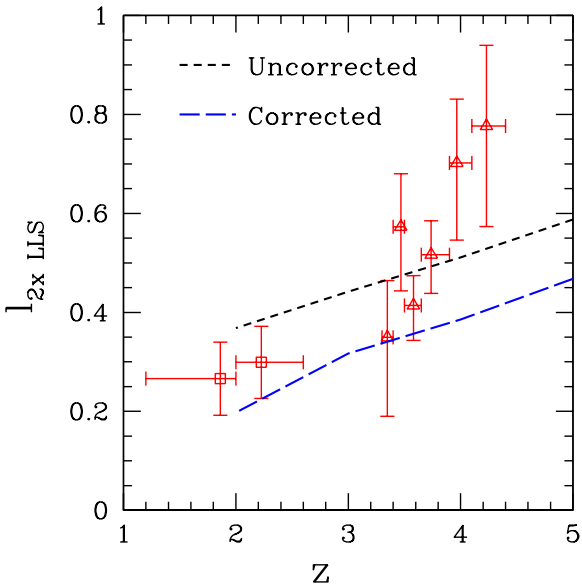}
\caption{Incidence of systems with $\NHI > 10^{17.5}\pcms$ in simulations and observations
as a function of redshift.
The short-dashed black curve shows the basic estimate from counting the number of absorbers
in the simulation and dividing by the absorption length of the simulation volume.
The long-dashed blue curve shows the result of correcting for the halo mass function.
The data are from two surveys: the squares are from \protect\cite{omeara_et_al_2012} and the triangles
are from \protect\cite{prochaska_et_al_2010}.}
\label{fig:dNdX}
\end{figure}

Due to the difficulty in directly measuring the column density of LLSs, the frequency of
LLSs per unit absorption length is the natural quantity to compare against observations.
I have computed this quantity using two approaches and plotted the result in
\Figref{fig:dNdX}. First, I counted the number of LLSs above $\NHI > 10^{17.5}\pcms$
along all of the sightlines in the simulation, and then divided by the absorption length
in the simulation:
\< l_{\rm \tau > 2} = \frac{\Delta \mathcal{N}_{\tau >2}}{\Delta X} .\>
The result of this simple approach is shown in \Figref{fig:dNdX} and is consistent with
observations although it has a somewhat different evolution in redshift.

In the second approach, I attempted to account for the bias inherent in a zoom-in
simulation by rescaling the contribution from each halo mass bin. Since the zoom-in
regions are selected to have a Milky Way progenitor, the mass function in these regions
will be biased as a random volume of this size would have fewer massive galaxies. One way
to account for this is to identify each LLS with its host halo, compute the mean
cross-section in each halo mass range, $\overline{\sigma}_{\tau > 2}(M_i,z)$, and then
compute the quantity
\< l_{\tau > 2} = \frac{c}{H_0} \sum_i \overline{\sigma}_{\tau>2}(M_i,z)
\overline{n}(M_i,z), \label{eq:discrete_dNdX}\>
where
\< \overline{n}(M_i,z) = \int_{M_i}^{M_i+\Delta M} n(M,z)\der M ,\label{eq:define_nbar}\>
and $n(M,z)$ is the true halo mass function. As long as the cross-section of individual
halos is correctly modeled, this discretized version of \Eqref{eq:dNdX_halo_mf} will
partially correct for the bias of the zoom-in simulation. Note that I have restricted
this sum to be over resolved halos with $M > 10^{8}h^{-1}\Msun$ (corresponding to
$\approx$1000 particles) below which we cannot model the cross-section and that I used
the halo mass function from \cite{sheth_tormen_2002} as the true halo mass function. Also
note that this sum is only covers the mass range of halos within the simulation but due
to the rapidly falling halo mass function and the relatively constant LLS covering
fraction which we will discuss in \Secref{sec:LLS_and_Halos}, the inclusion of higher
mass halos should not significantly change this result. The corrected incidence frequency
is plotted in \Figref{fig:dNdX}. It is lower than the basic counting result since it
lowers the contribution from more massive halos. While the simulated incidence frequency
is consistent with the observations until $z\sim 3.5$, there is significant deviation at
higher redshift. This is likely due to the zoom-in simulations used in this work which
cannot capture the contribution from the filamentary cosmic-web at high-redshifts. The
mean cross-section computed in the simulation can be found in  \Figref{fig:xsec_vs_mvir}
and will be discussed in more detail in \Secref{sec:LLS_and_Halos}.

This technique also relies on the properties of the galaxies in the zoom-in region being
representative of the properties of average galaxies in the universe. While this bias
cannot be addressed with individual zoom-in regions, simulations with fixed-resolution
\citep[i.e.][]{rahmati_et_al_2013_b} give similar results for the cumulative distribution
function (CDF) of LLSs with respect to halo mass, indicating that the assumption is a
reasonable one. In Figure 6 of \cite{rahmati_et_al_2013_b}, the CDF shows a similar
behavior to what is found in \Figref{fig:total_xsection} of this work with $\sim 75\%,
\sim 15\%,$ and $\sim 10\%$ of LLSs arising in halos with masses in the range $M <
10^{10} M_\odot$, $10^{10} M_\odot < M < 10^{11} M_\odot$, and $M > 10^{11} M_\odot$,
respectively at $z=3$. In this work we find $\sim 71\%$, $\sim 22\%,$ and $\sim 7\%$ of
LLSs arising in halos with the same mass range.

\subsection{Evolution of the HI Column Density Distribution}\label{sec:cdd_evolve}

\begin{figure}
\centering
\includegraphics[width=8cm]{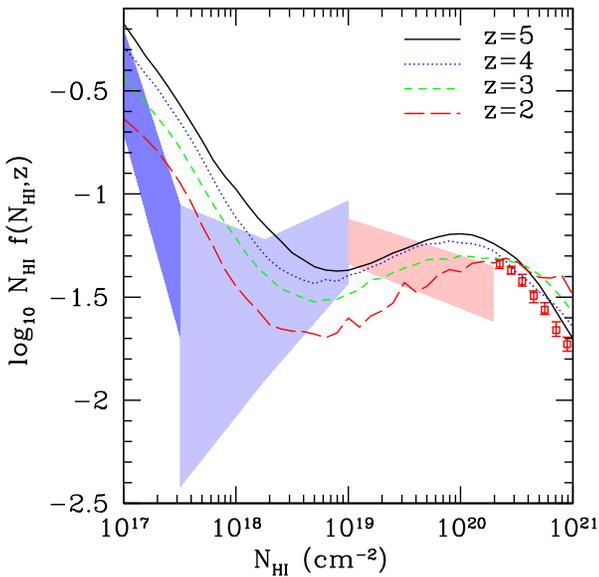}
\caption{HI column density distribution compared to observations centered around $z\approx 2.4$.
Since the column density distribution is fairly steep, I plot $\log_{10} \NHI f(\NHI,z)$ so that the features
are more salient. The light blue shaded region comes from constraints on $l_{\tau > 2}$ from \protect\cite{omeara_et_al_2012}.
The dark blue shaded region comes from constraints on the slope of the column density distribution
in the range $\NHI \in 10^{16.9}-10^{17.5} \pcms$ from the constraints on $l_{\tau > 2}$,
$l_{\tau > 1}$, and $l_{\tau > 0.5}$ \protect\citep{omeara_et_al_2012}. The light red region comes from \protect\cite{omeara_et_al_2007}.
The red squares come from \protect\cite{noterdaeme_et_al_2012}. Note that the column density
distribution from the simulations has not been re-scaled in any way. Since all of the observations
are centered around $z\approx 2.5$, they should be compared with the $z=2$ and $z=3$ column
density distribution.} \label{fig:cdd}
\end{figure}

In order to compute the column density distribution, I count the number of absorbers in
each HI column density bin, and divide by the total absorption length in the simulation:
\< f(\NHI) = \frac{\Delta \mathcal{N}(\NHI)}{\Delta \NHI\Delta X}. \>
In \Figref{fig:cdd}, I compare the HI column density distribution for LLSs in simulations
to observations. Since the column density distribution is quite steep over this range, I
have plotted the quantity $\log_{10} \NHI f(\NHI,z)$ in order to aid comparison. The HI
column density distribution in simulations has a qualitatively similar structure to the
observed HI column density distribution. The column density distribution is steep at low
$\NHI$ and then flattens out when self-shielding becomes important as I will discuss
further in \Secref{sec:individual_LLS}. Once the gas becomes sufficiently neutral, the
column density distribution steepens once again. This structure has been seen in many of
the recent simulations of Ly$\alpha$ absorbers
\citep[i.e.][]{mcquinn_et_al_2011,fumagalli_et_al_2011,altay_et_al_2011,rahmati_et_al_2012}.
In the observations, the flattening of the HI column density distribution is poorly
constrained since it occurs on the flat portion of the curve of growth where there are
only integral constraints on the HI column density distribution.

Interestingly, \Figref{fig:cdd} indicates that the HI column density distribution remains
relatively flat over a larger range than seen in the observations. A similar shape was
found in \cite{mcquinn_et_al_2011}. Note that since the quantity being plotted is
proportional to the number of absorbers per logarithmic $\NHI$ bin, \Figref{fig:cdd}
implies that there are more systems per logarithmic interval at $\NHI = 10^{20}\pcms$
than at $\NHI = 10^{19}\pcms$. A similar inversion is seen in the data although at
slightly lower column density. I will discuss the location of this turnover further in
\secref{sec:individual_LLS}.

From \Figref{fig:cdd}, it is apparent that the shape of the HI column density
distribution undergoes little evolution between $z=2$ and $z=5$, although there is a
slight flattening at low column densities and low redshift. This lack of evolution agrees
with the previous results found by \cite{fumagalli_et_al_2011} and
\cite{rahmati_et_al_2012}. Note that this work finds slightly less evolution in the
column density distribution from $z=5$ to $z=3$ than is found in
\cite{rahmati_et_al_2012}. This difference is likely due to the same reason that I
underpredict the incidence of LLS in \Figref{fig:dNdX}, the zoom-in simulations in this
work do not capture the large-scale filaments at high redshift.

\section{LLSs and Their Host Halos}\label{sec:LLS_and_Halos}

While these observations provide relatively unbiased statistics of the incidence of LLSs,
individual lines of sight cannot easily be used to study the halos in which LLSs reside.
Previous theoretical work has attempted to identify the host halos of these systems. Much
of the early work that explored the halo mass range lacked the mass resolution to study
absorbers in low-mass halos and extrapolated their properties from those of more massive
halos \citep[i.e.][]{katz_et_al_1996,abel_mo_1998,gardner_et_al_2001}. Making use of
simulations with better mass resolution, \cite{kohler_gnedin_2007} found that LLSs are
associated with a large range of halo masses but that low-mass halos do not dominate the
total cross-section of LLSs. More recent studies with similar resolution to this work
found that while LLSs are associated with a large range of halo masses, there is a
correlation between $\NHI$ and halo mass with lower column density systems more likely to
be found near lower mass halos
\citep[e.g.][]{van_de_voort_et_al_2012,rahmati_et_al_2013_b}. Using simulations with even
better mass resolution, as well as additional physics, I will now explore the relation
between LLSs and their host halos.

\subsection{LLS Cross-Section versus Halo Mass}

A simple statistic to consider is the mean LLS cross-section as a function of halo mass.
Some previous studies connect LLS and galaxies based on their projected separation. This
choice is similar to what is done in observational studies which most likely is the main
motivation for doing so in theoretical studies which aim to compare their results against
observations, \citep[e.g.][]{fumagalli_et_al_2011}. However, this can potentially lead to
unphysical correlations when the gas is near multiple halos in projection. In this work,
the nearest halo is instead determined by associating a given line of sight with the halo
closest to the maximum density point along the line of sight. By associating the LLS with
the nearest halo in 3-d space, the resulting cross-section should more accurately
represent the gas residing in that halo. The cross-section for each halo is computed in
each cartesian direction and then averaged.

\begin{figure}
\centering
\includegraphics[width=80mm]{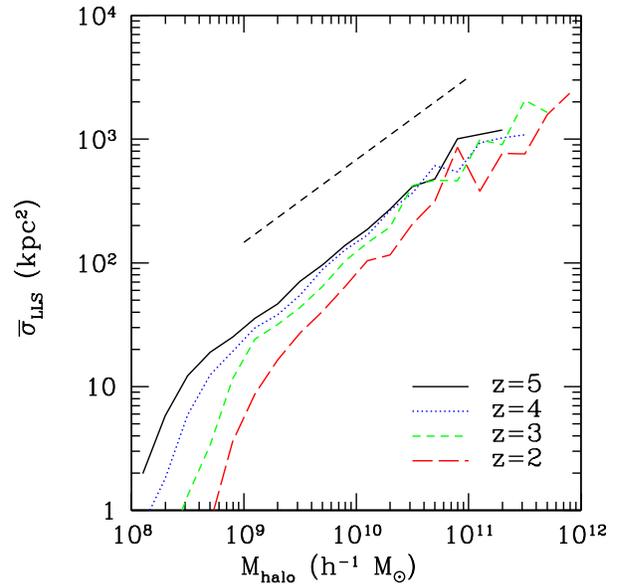}
\caption{Mean LLS cross-section versus the mass of the closest halo at different redshifts.
The black dashed line
is for reference and has a logarithmic slope of $\frac{2}{3}$. The curves have a similar
slope to this line, indicating that $\sigma_{\rm LLS} \propto r_{\rm vir}^2$. There is a
clear evolution in redshift with halos of a given mass having a smaller LLS cross-section
at lower redshifts. In addition, there is a cutoff at low mass which increases with decreasing redshift.}
\label{fig:xsec_vs_mvir}
\end{figure}

In \Figref{fig:xsec_vs_mvir}, I plot this mean cross-section for systems within a virial
radius of the host halo at four different redshifts. For reference, I also include a line
with a logarithmic slope of $\frac{2}{3}$. The average cross-sections have a similar
slope to this line, indicating that $\sigma_{\rm LLS} \propto r_{\rm vir}^2$ over a wide
range of halo masses. This implies that the halos have a fairly constant covering
fraction for LLSs within their virial radii. This covering fraction (both its magnitude
and mass independence) is similar to the values reported in \cite{fumagalli_et_al_2014}
with a $\sim 15\%$ covering fraction at $z=2$ and a $\sim 20\%$ covering fraction at
$z=3$ within the virial radius, in agreement with Figure 2 of their work. Given that
strong feedback is known to increase the LLS covering fractions
\citep[e.g.][]{faucher_et_al_2015,rahmati_et_al_2015} it is likely that these LLS
covering fractions are lower limits. The average cross-section also has a sharp drop-off
below a characteristic mass which I will discuss further below. The average cross-section
evolves with redshift in two ways. First, there is a decrease in the mean cross-section
at a given mass as the redshift decreases. Second, the characteristic mass below which
the cross-section drops-off increases with redshift. Note that if the LLS is instead
associated with the nearest and most massive halo within a projected virial radius, the
low mass halos, $M<10^9 {\rm h}^{-1} M_\odot$, will have a slightly lower cross-section
since some of the gas which belongs to them gets associated with a larger halo instead.

\begin{figure}
\centering
\includegraphics[width=80mm]{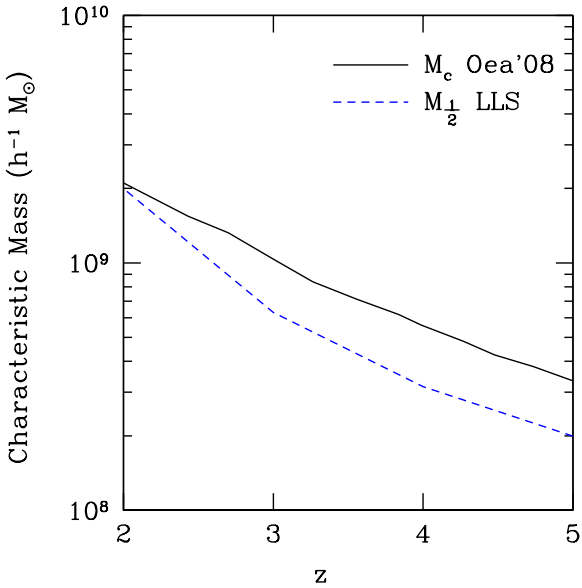}
\caption{Characteristic mass scale for halos to retain half their gas. The solid black curve
shows the characteristic mass from \protect\cite{okamoto_et_al_2008}, $M_c$, at which halos retain half
of their baryonic mass. The dashed blue curve shows the characteristic mass, $M_\frac{1}{2}$, at which the
covering fraction within the virial radius drops to half of the asymptotic value, as
described in the text.} \label{fig:okamoto_mass}
\end{figure}

This characteristic mass and its evolution can be interpreted in terms of the
photoionization of halos due to the UVB, a process described in
\cite{hoeft_et_al_2006,okamoto_et_al_2008}. In \cite{okamoto_et_al_2008}, the authors
studied the baryon fraction of halos as a function of halo mass and redshift. They found
that there is a characteristic mass which evolves with redshift at which the halos retain
half of the universal baryon fraction. Below this mass, the halos are unable to retain
their gas due to photoheating from the UVB. Note that the reference simulation used in
that work had a similar mass resolution ($2.2 \times 10^{5}h^{-1}\Msun$) to the
simulations used in this work so the same effect should be seen. Instead of the baryonic
fraction, I use the LLS covering fraction within a virial radius:

\<f_{\rm LLS} = \frac{\sigma_{\rm LLS}}{\pi r_{\rm vir}^2}.\>
For large halos, this covering fraction asymptotes to a constant value which depends on
redshift (see \figref{fig:xsec_vs_mvir}). I then find the characteristic mass at which
the covering fraction drops to half of this asymptotic value, $M_{\frac{1}{2}}$. Below
this mass, the covering fraction falls rapidly. I compare the characteristic mass derived
from the LLSs covering fraction with the characteristic mass from
\cite{okamoto_et_al_2008} in \Figref{fig:okamoto_mass}. I find that they roughly agree
and have a similar evolution with redshift which suggests that the drop in the LLS
covering fraction is due to photoionization of low-mass halos. Note that this comparison
is only a qualitative one since the characteristic mass as derived from the baryonic
fraction is not expected to be the same as the characteristic mass as derived from the
LLS covering fraction.

\subsection{Contribution of Different Mass Halos to the LLS Population}

\begin{figure}
\centering
\includegraphics[width=80mm]{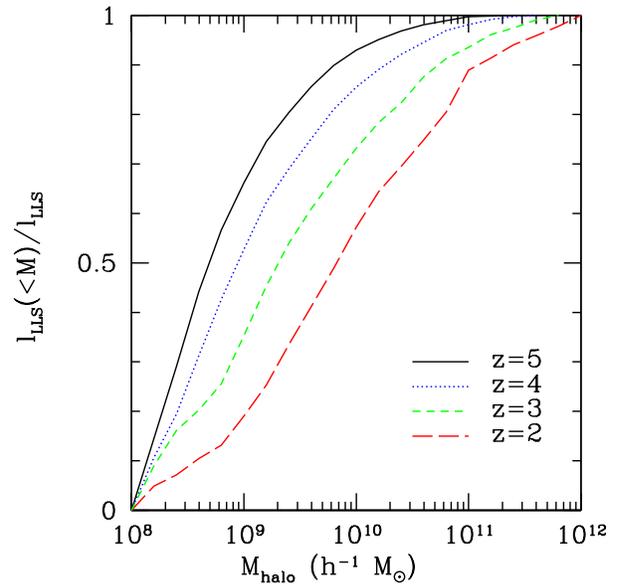}
\caption{Cumulative LLS incidence versus halo mass at different redshifts. Note that
the contribution from each mass range has been corrected by the halo mass function. While LLSs
arise in a variety of systems, most LLSs arise in low-mass halos with $M < 10^{10}h^{-1} M_\odot$.
This figure also shows a clear evolution in redshift: at later times, LLSs arise in more massive
halos.} \label{fig:total_xsection}
\end{figure}

Next, I compute how much each halo mass range contributes to the total LLS population.
The cumulative contribution to the LLS incidence for halos with mass less than $M$ is
given by
\< l_{\rm LLS} (< M) = \frac{c}{H_0} \sum_{M_i = M_{\rm min}}^M \overline{\sigma}_{\rm
LLS}(M_i,z) \overline{n}(M_i,z) , \label{eq:xsection_hmf}\>
where $M_{\rm min}$ is a minimum mass, given by $10^{8} h^{-1} \Msun$ in this work and
$\overline{n}(M_i,z)$ is defined as in \Eqref{eq:define_nbar}. This cumulative incidence
is plotted in \Figref{fig:total_xsection} where it has been normalized by the total
incidence. I find that a large range of halos contribute to the total LLS frequency.
Furthermore, I find that for redshifts between $z=2$ and $z=5$, low-mass halos with $M <
10^{10}h^{-1} \Msun$ contribute the majority of LLSs. While the contribution to the LLS
population from halos with $M < 10^{10}{\rm h}^{-1} \Msun$ has been studied before
\citep[i.e.][]{rahmati_et_al_2013_b}, the mass resolution used in this work allows us to
extend this to the population of LLSs residing in halos with $M < 10^9 {\rm h}^{-1}
\Msun$ which contribute $\sim 30\%$ of the total LLSs at $z=3$.

This mass range is especially interesting since H$_2$-based models of star formation
predict that these halos with $M< 10^{10}h^{-1} \Msun$ will have little star formation
and hence should be dark \citep{gnedin_kravtsov_2010,kuhlen_et_al_2013}. The results of
\Figref{fig:total_xsection} indicate that while these halos may be dark, they will
contribute the majority of systems seen in surveys of LLSs.

\subsection{Distance to the Nearest Halo}

\begin{figure}
\centering
\includegraphics[width=80mm]{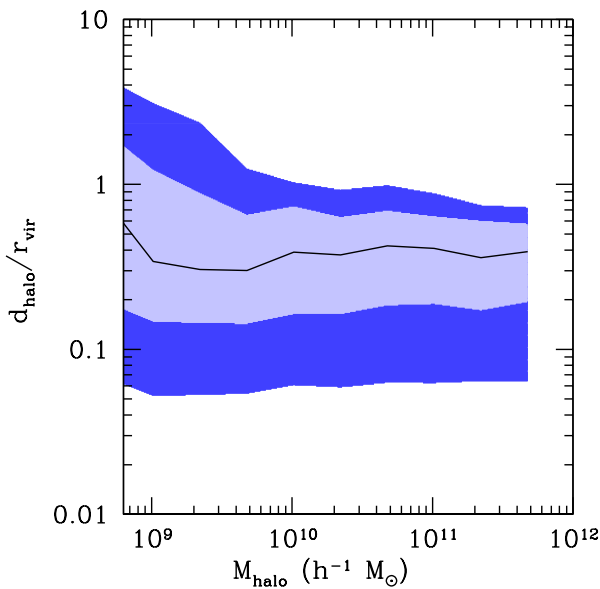}
\caption{3d distance to nearest halo, in units of the virial radius, as a function of halo mass.
Note that this plot is made from the $z=3$ snapshot which has a characteristic mass of $6.3\times 10^8 h^{-1}\Msun$.
The black curve is the median, the light blue (dark blue) band is the 1$\sigma$
(2$\sigma$) scatter around the median. The constancy of this ratio over a wide range of masses
indicates that the LLSs have a self-similar structure around their host halos where LLSs are found at
the same fraction of the virial radius.} \label{fig:dhalo_mvir}
\end{figure}

Now that I have explored the mass range of systems hosting LLSs, I will study the
distance from the LLSs to the nearest halo. In \cite{kohler_gnedin_2007}, the authors
showed that the distance to the nearest halo scaled like the virial radius, although this
relation had significant scatter due to the resolution of the simulation and the lack of
statistics. In \Figref{fig:dhalo_mvir}, I plot the median distance to the nearest halo in
units of the virial radius of the halo, as a function of halo mass. As expected from
\Figref{fig:xsec_vs_mvir}, there is a self-similar structure where LLSs can be found at a
constant fraction of the virial radius down to the cutoff mass. This plot is from the
$z=3$ snapshot which has a cutoff mass of $M_{\frac{1}{2}}=6.3 \times 10^8 h^{-1}\Msun$
(see \figref{fig:okamoto_mass}). Below this mass, the median distance to the nearest halo
is dominated by systems outside of the virial radius and hence the distance to the
nearest halo increases at low halo masses.

\begin{figure}
\centering
\includegraphics[width=80mm]{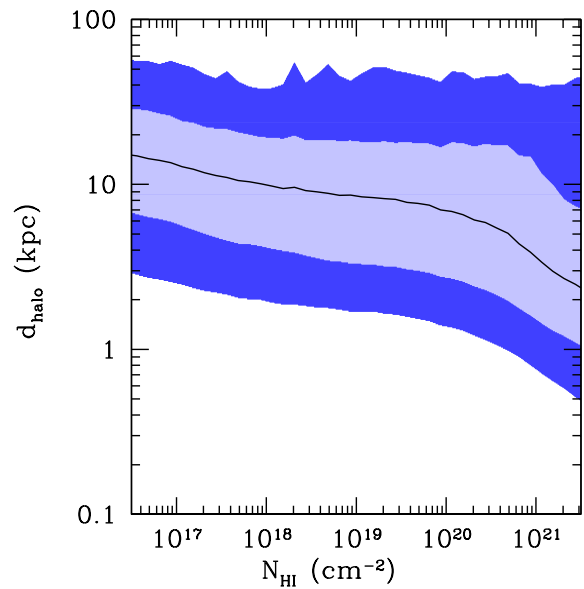}
\caption{3d distance to nearest halo in kpc as a function of the column density of the absorber.
The black curve is the median, the light blue (dark blue) band is the 1$\sigma$
(2$\sigma$) scatter around the median. The trend shows an anti-correlation between distance and $\NHI$
with stronger absorbers residing closer to their host halos.} \label{fig:dhalo_NHI}
\end{figure}

A related and important quantity is how the distance to the nearest halo depends on the
column density of the absorber. In \Figref{fig:dhalo_NHI} I plot the median distance to
the nearest halo as a function of $\NHI$. This shows an anti-correlation between the
distance to the halo and $\NHI$, i.e. stronger absorbers are closer to their host halo.
This trend is very similar to what was found in \cite{rahmati_et_al_2013_b}with a fairly
weak anti-correlation for LLSs which becomes stronger in the DLA regime\citep[see Figure
2 in][]{rahmati_et_al_2013_b}.

\section{Physical Properties of Individual LLSs} \label{sec:individual_LLS}

Now that I have explored the observed properties of LLSs, as well as the halos in which
these systems reside, I will study the physical nature of individual LLSs. LLSs span a
wide range of column densities: from $\NHI = 10^{17.2} \pcms$ to $\NHI=10^{20.3}\pcms$.
At the lower end of this range, the systems are mostly ionized and are believed to be in
photoionization equilibrium \citep{schaye_2001}. As the column density increases, these
systems become significantly self-shielded and become mostly neutral by the DLA
threshold. In this section I will explore this transition and test the model developed in
\cite{schaye_2001}.

\subsection{Analytical Model}

\cite{schaye_2001} developed a simple model to describe the properties of LLSs. At low
column densities, the gas is taken to be in photoionization equilibrium with the UVB,
i.e.
\< \Gamma \nhi = \beta_{\rm HII} n_{\rm e} n_{\rm HII} \label{eq:PIR} \>
where $\Gamma$ is the photoionization rate, $\beta_{\rm HII}$ is the recombination
coefficient, and $\nhi, n_{\rm HII}, n_{\rm e}$ are the number densities of HI, HII, and
electrons respectively. This relation can be used to solve for the HI fraction in terms
of the photoionization rate, recombination rate, and the hydrogen density. The
recombination rate is a function of the temperature which can be found in
\cite{draine_2011}.

In addition, \cite{schaye_2001} argues that the characteristic size of the absorber is
given by the Jeans length of the system:
\< L_J = t_{\rm ff} c_{\rm s} = 0.52\: {\rm kpc}\: \Big(\frac{n_{\rm H}}{1 {\rm
cm}^{-3}}\Big)^{-1/2} T_4^{1/2}, \label{eq:jeans_length}\>
where $T_4 = T/10^4 {\rm K}$ is the temperature of the gas and I have assumed that the
gas is at the universal baryon fraction. The assumptions of this model are spelled out in
detail in \cite{schaye_2001} and assume that the gas in hydrostatic and photoionization
equilibrium and that density distribution is uniform. Note that the temperature depends
weakly on $\NHI$ but is on the order of $10^4 {\rm K}$ for LLSs. The photoionization
equilibrium assumption breaks down as the system becomes significantly self-shielded and
at large $\NHI$, the gas becomes fully neutral. For systems at large $\NHI$, assuming
that the gas is fully neutral with a scale length given by the Jeans length gives the
correct asymptotic behavior but not the normalization.

\subsection{Characteristic Size}

The model developed in \cite{schaye_2001} assumes that the typical length of these
systems is given by the Jeans length. As a measure of the characteristic size of the
absorber, I take the length needed to get 90\% of the total HI absorption along a line of
sight. This mitigates the contribution of HI which is not associated with the LLS which
can lead to artificially large sizes. This scheme was used by \cite{prochaska_wolfe_1997}
where they faced a similar problem in measuring the velocity width from a metal-line
absorption profile. I implement this method by taking 500kpc lines of sight centered on
the absorber and determining the HI column density along this line of sight. I then find
the distance needed to get 45\% of the total $\NHI$. I have tested that this
characteristic length has converged by considering longer lines of sight (up to 1Mpc).

\begin{figure}
\centering
\includegraphics[width=8cm]{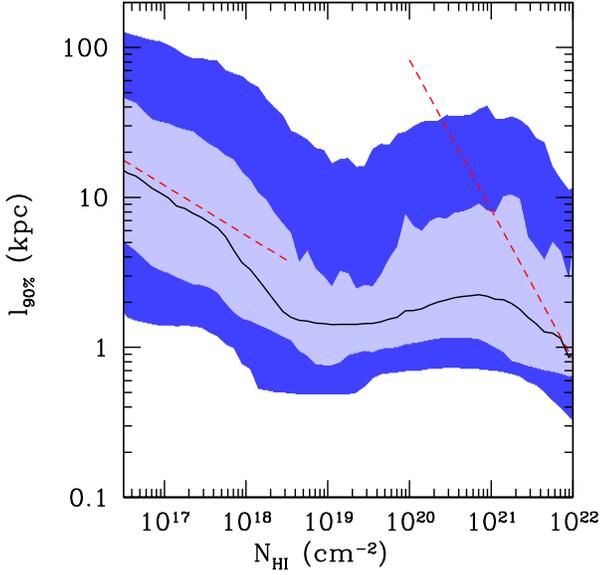}
\caption{Absorption length for 90\% of the absorption. The solid black curve is the median and the light blue (dark blue)
band is the 1$\sigma$ (2$\sigma$) scatter around the median. At low $\NHI$, the dashed red
line is the Jeans length assuming photoionization equilibrium and $T=1.5\times10^4$K - close to the average
temperature in the simulation at these $\NHI$. At high $\NHI$,
the dashed red line is the Jeans length assuming that the gas is fully neutral with an arbitrary
normalization to show that the model recovers the scaling behavior.}
\label{fig:abs_length}
\end{figure}

In \Figref{fig:abs_length} I plot the median characteristic length as a function of
$\NHI$ along with the model from \cite{schaye_2001}. For the low $\NHI$ systems, I have
over-plotted the Jeans length assuming photoionization equilibrium. For the high $\NHI$
systems, I over-plotted the Jeans length assuming the gas is fully neutral. At low
$\NHI$, I find that the model is very close to the median. Note that the model should not
be expected to give an exact quantitative match but rather describe the scaling and
trends of the simulation results. Most importantly, the model reproduces the scaling
behavior at low $\NHI$, $L_J \propto \NHI^{-1/3}$, which follows from combining
\Eqref{eq:PIR} and \Eqref{eq:jeans_length}, i.e. assuming that the gas is in
photoionization equilibrium with the UVB and in local hydrostatic equilibrium. This is a
good assumption for optically thin gas at high redshift and explains the relation between
column density and density seen in Ly$\alpha$ forest simulations
\citep[e.g.][]{dave_et_al_2010,mcquinn_et_al_2011,altay_et_al_2011,rahmati_et_al_2012}.

\subsection{Transition from Ionized to Neutral LLSs} \label{sec:shielding}

\begin{figure}
\centering
\includegraphics[width=8cm]{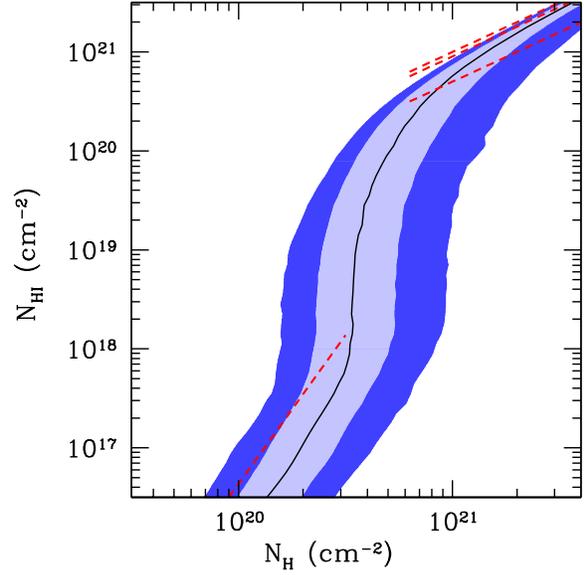}
\caption{$\NH$ versus $\NHI$ along 200 kpc lines of sight. The solid black curve is the median and the light blue (dark blue)
band is the 1$\sigma$ (2$\sigma$) scatter around the median. At low $\NH$ I have
assumed photoionization equilibrium. The dashed red line corresponds to $T=1.5\times10^4$K - the average
temperature at these $\NHI$ in the simulation. Although the model does not quantitatively
match the median, it does reproduce the scaling behavior of $\NHI \propto \NH^3$ which is
described in the text. At large $\NH$, the gas becomes neutral and asymptotically approaches $\NHI = \NH$ until molecular hydrogen effects and
ionization from local sources become important. To guide the eye, I have included curves with $\NHI = 0.5\NH$, $\NHI = 0.9\NH$, and $\NHI=\NH$ which
are lines of constant HI fraction. The median in the simulation is asymptoting to fully neutral gas.} \label{fig:NHI_vs_NH}
\end{figure}

As the HI column density increases from the threshold of a LLS up to a DLA, the systems
go from mostly ionized to neutral due to self-shielding. In \Figref{fig:NHI_vs_NH}, I
plot the median HI column density versus the total hydrogen column density along 200kpc
lines of sight centered on the absorber. As in the previous plots, these quantities are
computed along lines of sight through the box. Note that I have plotted the total $\NH$
on the $x$-axis to emphasize that $\NHI$ depends on the total $\NH$. Since the average HI
fraction along a line of sight is given by $\NHI/\NH$, this plot also shows how the HI
fraction depends on $\NH$.

At low column density, $\NHI < 10^{18} \pcms$, I have included the photoionization
equilibrium model with the UVB. The gas is taken to be highly ionized and in
photoionization with the UVB. The column densities are thus given by $\NHI = n_{\rm HI}
L_J \propto n_{\rm H}^{\frac{3}{2}}$ and $\NH = n_{\rm H}L_J \propto n_{\rm
H}^{\frac{1}{2}}$ at constant temperature. Although this model does not quantitatively
match the simulation result, it does reproduce the scaling behavior of $\NHI \propto
\NH^3$. The main reason for the discrepancy is that atomic hydrogen is more localized
that the total hydrogen since it must be self-shielded. As a result, for the 200 kpc line
of sight used in \Figref{fig:NHI_vs_NH}, $\NH$ gets a more substantial contribution from
material outside the Jeans length which offsets the relation to the right of the model at
low column densities. The quantity considered below, $\langle n_{\rm H} \rangle$, avoids
this problem and has a better match at low $\NHI$.

Above the threshold of $\NHI = 10^{18}\pcms$, there is a rapid increase in $\NHI$ for a
small increase in $\NH$ due to self-shielding of the gas. For the highest column density
systems, $\NHI
> 10^{20.3} \pcms$, the systems asymptote to fully atomic systems. To showcase this
asymptotic behavior, I have included 3 lines in \Figref{fig:NHI_vs_NH} with successively
higher neutral fractions. Note that at even higher column densities, molecular physics
becomes important and non-negligible H$_2$ fractions make $\NHI < \NH$.

\begin{figure}
\centering
\includegraphics[width=8cm]{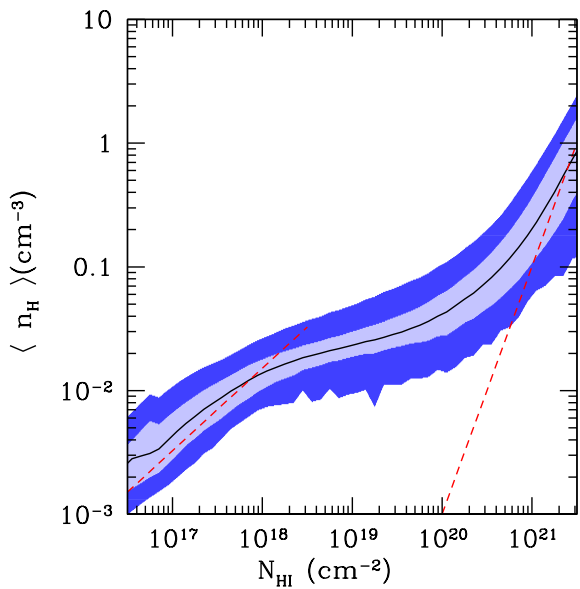}
\caption{$n_{\rm H}$ weighted by $n_\HI$ averaged along 200 kpc sightlines versus $\NHI$. The
solid black curve is the median and the light blue (dark blue)
band is the 1$\sigma$ (2$\sigma$) scatter around the median.
At low $\NHI$, the dashed red curve shows the prediction from the photoionization equilibrium model with $T=1.5\times 10^4$K which
reproduces the scaling behavior of the median. At large $\NHI$, the dashed red curves show the
prediction from fully neutral gas with an arbitrary normalization to show the model recovers the scaling behavior} \label{fig:nH_vs_NHI}
\end{figure}

A related plot found in other works
\citep[i.e.][]{mcquinn_et_al_2011,altay_et_al_2011,rahmati_et_al_2012} is the median gas
density versus $\NHI$. As in these works, I compute the integral of $\nh$ weighted by
$\nhi$:
\< \langle \nh \rangle = \frac{\int \nh \nhi dl}{\int \nhi dl} .\>
Since $\nhi$ is more sharply peaked than $\nh$ due to self-shielding, this effectively
selects the central part of the absorber. I show the median $\langle \nh \rangle$ in
\Figref{fig:nH_vs_NHI}. I find that the photoionization equilibrium model reproduces the
properties well at low $\NHI$. It matches the scaling behavior of $\langle \nh \rangle
\propto \NHI^{2/3}$ derived from \Eqref{eq:PIR} and \Eqref{eq:jeans_length}. Above $\NHI
= 10^{18}\pcms$, self-shielding becomes important and there is a large increase in $\NHI$
for a small increase in $\langle \nh \rangle$. At the highest $\NHI$, the gas is expected
to be fully neutral and the model from \cite{schaye_2001} predicts that $\langle \nh
\rangle \propto \NHI^2$. As in \Figref{fig:abs_length}, the median does not asymptote to
the model curve.

\begin{figure}
\centering
\includegraphics[width=8cm]{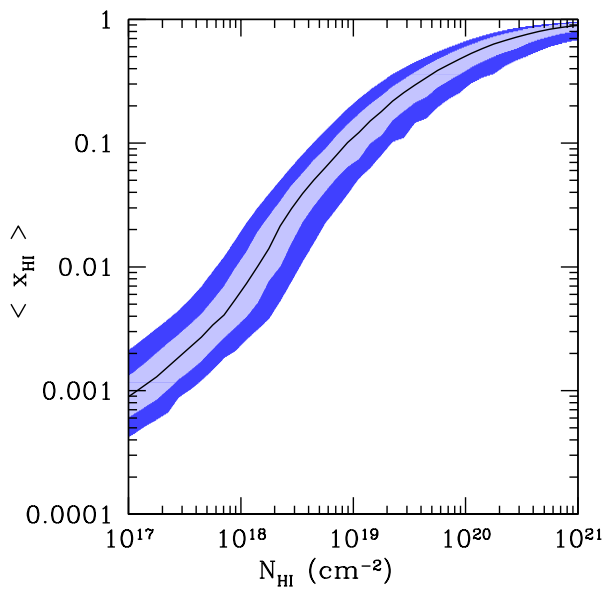}
\caption{$x_{\rm HI}$ weighted by $n_\HI$ averaged along 200 kpc sightlines versus $\NHI$. The
solid black curve is the median and the light blue (dark blue)
band is the 1$\sigma$ (2$\sigma$) scatter around the median.} \label{fig:xHI_vs_NHI}
\end{figure}

For ease in comparison with other work, I also include a related quantity which is the
$n_{\rm HI}$ weighted $x_{\rm HI}$ fraction in \Figref{fig:xHI_vs_NHI}
\citep{mcquinn_et_al_2011,altay_et_al_2011}. The comparison between the results of those
works and this work is discussed in \Secref{sec:other_works}.

\subsection{Effect of Self-Shielding on the Column Density Distribution}
\label{sec:cdd_fNH_vs_fNHI}

\begin{figure}
\centering
\includegraphics[width=8cm]{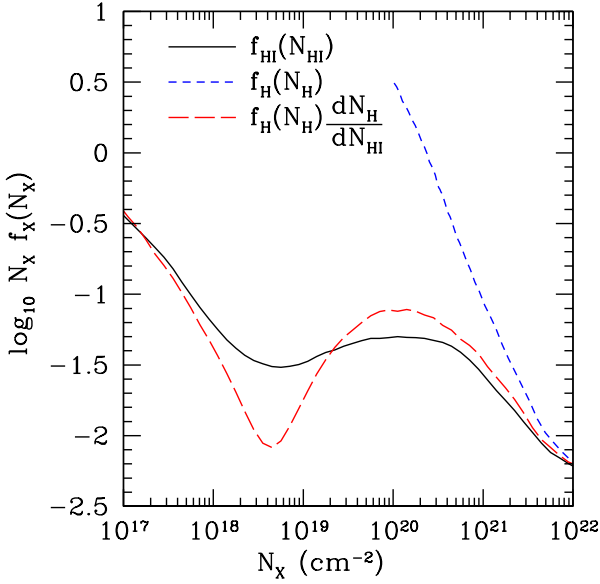}
\caption{HI column density distribution and H column density distribution. The black solid
curve shows the HI column density distribution as computed from the simulation. The blue, short-dashed, curve shows
the total hydrogen column density distribution as computed in the simulation. Finally, the red long-dashed curve shows the result of taking the median
profile in \protect\Figref{fig:NHI_vs_NH} to compute $\frac{d\NH}{d\NHI}$ and then computing the HI column density
distribution using \protect\Eqref{eq:fNHI_fNH}. Note that the median relation between $\NHI$ and $\NH$ was smoothed
over in order to reduce the noise in the derivative.} \label{fig:fNH_vs_fNHI}
\end{figure}

In \Secref{sec:cdd}, we saw that the HI column density distribution has a flattening at
$\NHI \sim 10^{18}\pcms$ which has been attributed to self-shielding
\citep{mcquinn_et_al_2011,altay_et_al_2011,rahmati_et_al_2012}. A priori it is unclear
that this flattening is only due to self-shielding and not due to some feature in the
total hydrogen column density distribution. This can be checked by comparing the HI
column density distribution, $f_\HI(\NHI)$, and the total hydrogen column density
distribution $f_{\rm H}(\NH)$, where I have included additional subscripts to emphasize
that they are different distributions. These two distributions are related by
\< f_\HI(\NHI) = f_{\rm H}(\NH) \frac{d\NH}{d\NHI}. \label{eq:fNHI_fNH}\>
The relation between $\NHI$ and $\NH$ is shown in \Figref{fig:NHI_vs_NH}. Using the
median of this relation, $\frac{d\NH}{d\NHI}$ can be computed. Furthermore, $f_{\rm
H}(\NH)$ can be computed in the simulation and then \Eqref{eq:fNHI_fNH} can be used to
compute $f_\HI(\NHI)$. The result of this procedure is shown in \Figref{fig:fNH_vs_fNHI}.
$f(\NH)$ is a power-law over the range in which the transition between ionized and
self-shielding occurs. Therefore, these simulations show that the feature at $\NHI \sim
10^{18}\pcms$ is a signature of self-shielding and not the distribution of the total
hydrogen at the corresponding column density.

\subsection{Photoionization Rate}

\begin{figure}
\centering
\includegraphics[width=8cm]{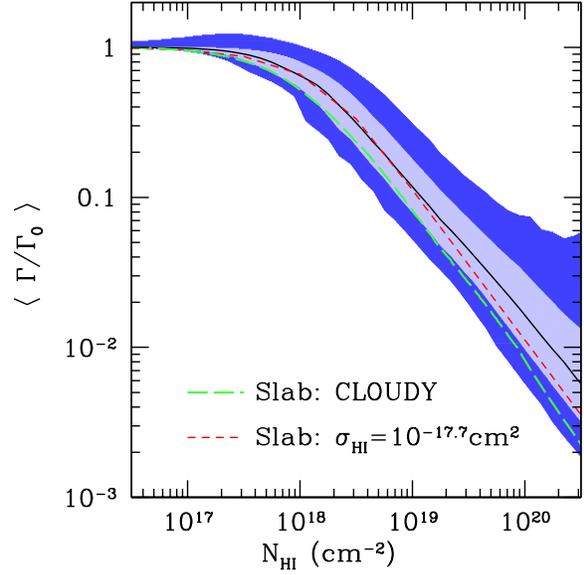}
\caption{Median photoionization rate versus $\NHI$. The photoionization rate is averaged along
sightlines and weighted by the HI density. The solid black curve is the median and the light blue (dark blue)
band is the 1$\sigma$ (2$\sigma$) scatter around the median. The short-dashed red curve is the model of the photoionization
rate from \protect\Eqref{eq:PI_simplified} which assumes a mono-chromatic UVB with $\sigma_{\rm HI} = 10^{-17.7}{\rm cm}^2$ and has a column density of $\NHI/2$.
The long-dashed green line comes from computing the average photoionization rate, \protect\Eqref{eq:PI_exact}, of a slab with
column density $\NHI/2$ illuminated by the UVB and was done with CLOUDY. These models are discussed
further in the text.}
\label{fig:PI_rate}
\end{figure}

In the limit where we can neglect radiative recombination and local sources of radiation,
the photoionization rate of LLSs directly measures the self-shielding of the LLS against
the UVB. Since the distance of an absorber from its host galaxy is anti-correlated with
its HI column density, as shown in \Figref{fig:dhalo_NHI}, low $\NHI$ systems will not be
significantly affected by the local radiation from their host halo. The decrease in the
photoionization rate in a LLS allows us to measure the effective shielding of the LLS
against the UVB. In \Figref{fig:PI_rate}, I plot the photoionization rate averaged along
lines of sight through the LLS, weighted by $\nhi$:
\< \langle \Gamma \rangle = \frac{\int \Gamma(l) n_{\HI} dl}{\int n_{\HI} dl}
.\label{eq:PI_exact}\>
If only the contribution from the UVB is considered, this integral can be solved for a
monochromatic UVB. In this limit, the differential optical depth can be written as $d\tau
= \nhi \sigma_{\rm HI} dl$ and get
\< \langle \Gamma \rangle = \frac{\int \Gamma_0 e^{-\tau} \sigma_{\rm HI}^{-1}
d\tau}{\NHI}, \>
where $\Gamma(\tau) = \Gamma_0 e^{-\tau}$ and $\sigma_{\rm HI}$ is independent of $\tau$.
This then gives
\< \langle \Gamma \rangle = \Gamma_0 \frac{1-e^{-\NHI \sigma_{\rm HI}}}{\NHI \sigma_{\rm
HI}} .\label{eq:PI_simplified}\>
In \Figref{fig:PI_rate}, I include this model for a slab with column density $\NHI/2$ and
find that a value of $\sigma = 10^{-17.7}{\rm cm}^2$ provides a fairly good fit at low
$\NHI$ although it does not match the slope at large $\NHI$. I use a column density of
$\NHI/2$ since the LLS is illuminated on all sides by the UVB and this model assumes that
the LLS is being illuminated from one direction. The difference between this model and
the median photoionization rate in the simulation for $\NHI > 10^{19} {\rm cm}^{-2}$ is
due to the increasingly important effects of radiative recombination and local radiation
as $\NHI$ increases
\citep[i.e.][]{miralda-escude_2005,schaye_2006,rahmati_et_al_2013,rahmati_et_al_2013_b}.
However, this effect is unimportant for determining the effective shielding of the LLS
which is determined at lower column densities.

I also include a model for the average photoionization rate for a slab with column
density $\NHI/2$ illuminated on one side by the UVB using CLOUDY v13.01
\citep{ferland_2013}. For this model, I set up a slab with a plane-parallel geometry,
irradiated by the Haardt-Madau background given in \cite{haardt_madau_2001}, with
appropriate helium and metal abundances. I varied the hydrogen density ($n_{\rm H} \in
[10^{-3},10^{-1}] {\rm cm}^{-3}$) and the metallicity ($Z/Z_\odot \in [10^{-3},10^{-1}]$)
and computed the HI photoionization rate as a function of HI column density through the
slab. I found that this relationship was robust and did not depend on the HI density or
metallicity. This result gives the long-dashed green curve in \Figref{fig:PI_rate} which
can be compared to the photoionization rate in actual simulations. This model has an
effective cross-section of $\sigma_{\rm HI} = 10^{-17.6}{\rm cm}^{2}$ at low column
densities. Interestingly, this model does not quantitatively match the absorption seen in
the simulation. This discrepancy is due to the anisotropy of the LLS which I will discuss
in the next section.

\section{Anisotropic Shielding of LLSs}\label{sec:anisotropy}

\begin{figure}
\centering
\includegraphics[width=8cm]{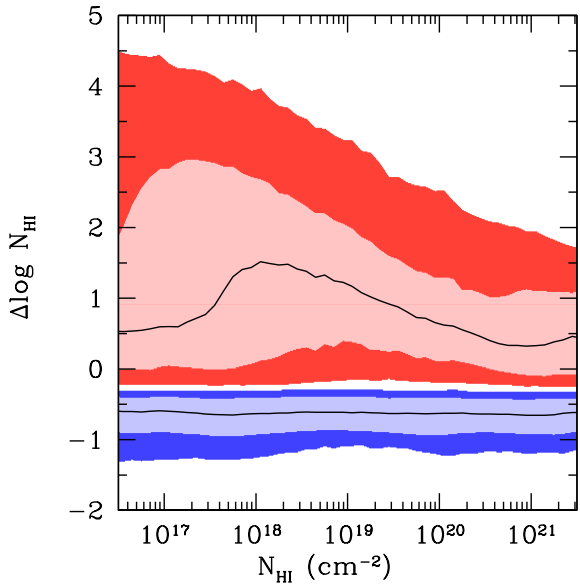}
\caption{Comparison of the difference between the column density along a line of sight and the column density along different directions originating from the absorber.
The red shaded (upper) region shows the 1 and 2$\sigma$ scatter of the difference between the maximum $\NHI$ and the blue
shaded (lower) region shows the 1 and 2$\sigma$ scatter of the minimum $\NHI$
along the 6 cartesian directions originating at the center of the absorber. The black curves in the center of each region show the median.} Note that the
column density on the $x$ axis is the column density through the entire system. This was
chosen to highlight the difference between the observed $\NHI$ of a system along a random
line of sight, and the characteristic minimum/maximum $\NHI$ between the center of the absorber
and the UVB.
\label{fig:NHI_directional}
\end{figure}

In the previous section, I tested the model developed in \cite{schaye_2001} and found
that it successfully reproduced may of the properties of LLSs. In this model, LLSs are
characterized by a single column density and the self-shielding of the absorber depends
on this quantity. However, for a non-spherical absorber the column density will depend on
the angular direction. To test the importance of this column density variation, I first
identified the centers of LLS by finding the maximum density along a line of sight.
Around this maximum, I then compute the column density along the 6 cartesian directions
originating from this point to determine the HI column density in these 6 directions. In
\Figref{fig:NHI_directional}, I show the column density along the original line of sight,
$\NHI$, versus the difference between $\NHI$ and the minimum/maximum column density in
the other 6 cartesian directions.

\Figref{fig:NHI_directional} shows that if a random line of sight in the system has a
column density of $\NHI$, on average there will be a line of sight originating from the
center of that system with a column density 0.6-0.7 dex lower, approximately $\NHI/4$. As
a result, systems will be more ionized than naively expected from the column density in a
single direction. This result is important for understanding the column density
distribution (\figref{fig:cdd}), as well as the relationship between $\NHI$ and $\NH$
(\figref{fig:NHI_vs_NH}).

\begin{figure}
\centering
\includegraphics[width=8cm]{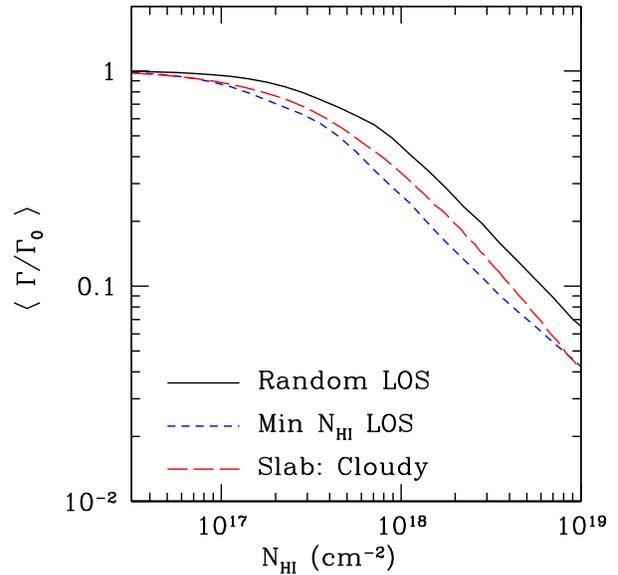}
\caption{Photoionization rate versus $\NHI$ in two different directions. The solid black curve
is the median of the photoionization along a a specific cartesian direction and hence along
an effectively random direction. As described in the text, the short-dashed blue curve is the median along the direction with the minimum $\NHI$ originating
from the center of the LLS. The long-dashed red curve is the photoionization
rate from CLOUDY assuming the Haardt-Madau background at $z=3$ \citep{haardt_madau_2001}.}
\label{fig:PI_rate_min}
\end{figure}

In \Figref{fig:PI_rate_min}, I compare the average photoionization rate along a cartesian
direction with the average rate along the direction with the lowest $\NHI$. Since the
absorbers are randomly oriented with respect to the box, this cartesian direction probes
an effectively random direction with respect to the absorber. The average photoionization
along this direction is given by the black solid curve. Fitting this curve using
\Eqref{eq:PI_simplified} gives an effective cross-section of $\sigma_{\rm HI} =
10^{-17.7} {\rm cm}^2$ at low column densities. The second direction is the direction
originating from the center of the LLS with the lowest $\NHI$. The short-dashed blue
curve shows the average photoionization rate versus column density along this direction.
I also include a slab model using the UVB in the simulation. This is done using CLOUDY as
I described in \Secref{sec:individual_LLS} and is given by the long-dashed red curve.

By comparing the curves in \Figref{fig:PI_rate_min}, I find that the photoionization rate
from the slab model in CLOUDY falls between the rate along a random direction and the
rate along the minimum direction in the simulation. This comparison is useful since it
shows that if one takes a random line of sight through a LLS, the gas along this line of
sight is less shielded than one would expect from the HI column density. This makes sense
since, on average, there will be a line of sight to the UVB with a significantly lower
column density (see \figref{fig:NHI_directional}) allowing for more photoionization than
naively expected. Likewise, for gas along the direction with the lowest column density,
there will be lines of sight with higher column densities which will result in a lower
photoionization rate than expected.

\section{Effective Shielding of LLSs} \label{sec:effective_shielding}

Putting together the results of \Secref{sec:individual_LLS} and \Secref{sec:anisotropy},
I find that the self-shielding of LLSs against the UVB is less than naively expected.
Given a LLS with column density $\NHI$, one would expect that this system is shielded by
an optical depth of $\tau = \NHI \sigma_{\rm HI}$, where $\sigma_{\rm HI}$ is an
effective cross-section of HI to the UVB. Since the self-shielding of LLSs is known to
flatten the column density distribution \citep[i.e.][or Section 5.5 of this
work]{altay_et_al_2011,mcquinn_et_al_2011,rahmati_et_al_2012}, it is important to
understand at what column density one should expect self-shielding to become important.

There are three effects which lower the amount of shielding. First, as I discussed in
\Secref{sec:individual_LLS}, since a LLS is bathed in the UVB from all sides, a system
with a column density of $\NHI$ is effectively only shielded by a column density of
$\NHI/2$. Second, the UVB is not monochromatic but has a spectrum which extends to high
energies. Since the cross-section of HI decreases with increasing energy, these photons
can penetrate deeper into the cloud and lower the effective cross-section of LLS to the
UVB. As I showed in \Figref{fig:PI_rate_min}, the effective cross-section against the UVB
at $z=3$ is $\sigma_{\rm HI} \approx 10^{-17.6}\pcms$, 0.4 dex lower than the
cross-section at the Lyman limit. Lastly, I investigated the effect of the anisotropy of
the LLS in \Secref{sec:anisotropy} and found that, on average, a LLS with a column
density of $\NHI$ will have a line of sight with column density $\NHI/4$ from the center
of the LLS to the UVB, i.e. half of what one would expect if the LLS was isotropic. This
anisotropy means that an average LLS will be less shielded than expected from the column
density. In \Figref{fig:PI_rate_min}, I found that this results in a $0.1-0.2$ dex
decrease in the optical depth as compared to a uniform slab.

Altogether, these three effects mean than a LLS need to have a column density of $\NHI
\sim 10^{18}\pcms$ in order to have an optical depth of unity. Since the flattening of
the column density distribution is due to this self-shielding, this means that we should
expect the column density distribution to start flattening around $\NHI \sim
10^{18}\pcms$, as I find in \Figref{fig:cdd}. In addition, the onset of self-shielding
can clearly be seen in the relation between $\NHI$ and $\NH$ in \Figref{fig:NHI_vs_NH}.
Note that the effective cross-section of HI also depends weakly on the redshift of the
LLS since the spectral shape of the UVB changes slowly with redshift.

\section{Comparison with Previous Work}\label{sec:other_works}

Both LLSs and DLAs have received significant attention in the literature and attempts are
now being made to quantitatively match observations. In this section, I will compare the
results in this work to papers which have made a similar attempt to understand the
properties of LLSs.

\cite{kohler_gnedin_2007} studied LLSs using simulations which had lower spatial and mass
resolution than the simulations in this work. They found many of the same trends found
here although they were limited on the low-mass end. They also studied the properties of
absorbers as a function of their parent halo and found that LLSs reside in halos with a
large range of masses and concluded that the majority of LLSs do not reside in very
low-mass halos. As in this work, they found that LLSs remain ionized up to fairly high
column densities, $\NHI = 10^{20}\pcms$. Despite including many of the physical
mechanisms needed to model the ionization state of the gas, their column density
distribution did not show any signs of self-shielding around $\NHI = 10^{18}\pcms$.

\cite{mcquinn_et_al_2011} studied LLSs using simulations with a similar simulation volume
as this work. They found a similar HI column density distribution as was found in this
work, with significant flattening due to self-shielding starting a little above $\NHI
\sim 10^{18}\pcms$. They also made comparisons to the model in \cite{schaye_2001} and
found that this model had a qualitative agreement with their results. Just as in this
work, they found that LLSs remain ionized up to high column densities, as can be seen in
the middle panel of Figure 5 in \cite{mcquinn_et_al_2011} , where they have a $n_{\rm
HI}$ weighted neutral fraction of $\sim 0.1$ at $\NHI = 10^{19} \pcms$, consistent with
the neutral fraction reported in \Figref{fig:xHI_vs_NHI} of this work.

\cite{altay_et_al_2011} studied both LLSs and DLAs and found a nice agreement with
observed column density distribution over a wide range of $\NHI$ and find self-shielding
starts to flatten the HI column density distribution above $\NHI = 10^{18}\pcms$.
Interestingly, the LLSs in their simulations are significantly less ionized than in this
work or in \cite{mcquinn_et_al_2011}. The left panel of Figure 3 in that works shows that
the $n_{\rm HI}$ weighted neutral fraction at $\NHI = 10^{19} \pcms$ is approximately
-0.2 dex, as compared to the -1 dex reported in \cite{mcquinn_et_al_2011} and
\Figref{fig:xHI_vs_NHI} of this work. Despite this difference in the ionization fraction,
their relation between $\langle n_{\rm H} \rangle$ versus $\NHI$ is very similar to what
was found in this work in \Figref{fig:nH_vs_NHI}.

\cite{rahmati_et_al_2012} studied the redshift evolution of the column density
distribution and found a similar evolution as to \Figref{fig:cdd}. While the amplitude
decreases with decreasing redshift, they find that the column density distribution
becomes slightly shallower at lower redshifts and low column densities. As was discussed
in \Secref{sec:cdd_evolve} , the overall normalization of their HI column density
distribution evolves more than this work between $z=5$ and $z=3$. This is likely due to
the same reason this work had difficulty reproducing the frequency of LLSs at
high-redshift in \Figref{fig:dNdX}: since this work uses zoom-in simulations, it does not
capture the large-scale filamentary structure at high redshift.

\cite{rahmati_et_al_2013_b} discussed many of the same properties of LLSs as in this work
using fixed dark matter particle mass of $6.3 \times 10^{6}$ h$^{-1} M_\odot$ ,as opposed
to the zoom-in simulations used in this work. The comparison between this work and
\cite{rahmati_et_al_2013_b} provides a good test of the assumption that the zoom-in
region is not overly biased. The cumulative LLS incidence with respect to halo mass is
also computed in the top right panel of Figure 6 in\cite{rahmati_et_al_2013_b} and shows
that there is not a large contribution from halos above $10^{12} M_\odot$, a range which
is inaccessible with the zoom-in simulations used in this work. On the low-mass end, the
simulations show that there is a significant contribution from halos below a halo mass of
$10^{10} M_\odot$, in agreement with \Figref{fig:total_xsection} of this work.
\cite{rahmati_et_al_2013_b} also studied the impact parameter of LLSs and found similar
results to \Figref{fig:dhalo_NHI} an anti-correlation between $N_{\rm HI}$ and the
distance to the nearest halo.

\section{Summary and Conclusion} \label{sec:conclusion}

In this work, I have explored the properties of LLSs using cosmological zoom-in
simulations which include on the fly radiative transfer and have high mass resolution.
The simulations in this work reproduce the observed incidence frequency of LLSs as well
as the HI column density distribution, indicating that the simulations are effectively
modeling LLSs.

Using these simulations, I investigated the host halos of LLSs.  The high mass resolution
of these simulations allowed me to investigate the LLS content of halos down to $10^8
{\rm h}^{-1} M_\odot$. These results showed that halos have a nearly constant covering
fraction of LLSs within their virial radius over a wide range of halo masses, similar to
the results in \cite{fumagalli_et_al_2014}. However, it is important to note that the
simulations use in this work, as well as those in \cite{fumagalli_et_al_2014} use
inefficient feedback which leads an overproduction of stellar mass in the halos of
interest. As has been recently shown in \cite{faucher_et_al_2015} and
\cite{rahmati_et_al_2015}, including more efficient feedback which is needed to produce
realistic stellar masses also increases the covering fraction of LLSs and boosts it to
values significantly higher than what is found in this work and in
\cite{fumagalli_et_al_2014}. Efficient feedback will likely affect many of the properties
of LLSs and this will be investigated in future work.

In addition to this near-constant covering fraction, there is a cutoff at low halo masses
which increases as the redshift decreases. I argued that this evolution of the cutoff is
real since the simulations have the necessary mass resolution to adequately model these
halos and that the evolution can be explained by the photoionization of gas in the galaxy
due to the UVB. In addition, I found that between $z=2-5$, more than $50\%$ of LLSs
reside in halos with $M<10^{10} {\rm h}^{-1} M_\odot$. This is especially interesting
since H$_2$-based star formation models predict that these galaxies will be dark
\citep[i.e.][]{gnedin_kravtsov_2010,kuhlen_et_al_2013}. As a result, absorption line
studies of LLSs will be an important testing ground for simulations since they probe a
large reservoir of gas which will be difficult to detect with other means.

Next, I investigated the properties of individual LLSs. I tested a simple model from
\cite{schaye_2001} and found that it reproduced the characteristic size and HI fraction
of LLSs well for $\NHI < 10^{18} \pcms$. Above this threshold, the gas is no longer
optically thin and the model is no longer valid. However, in the DLA regime, the gas is
almost entirely neutral so the simple model is justified once again with a scale length
given by the Jeans length. Using the relation between $\NHI$ and $\NH$, I showed how
onset of self-shielding at $\NHI = 10^{18}\pcms$ is responsible for the flattening of the
HI column density distribution which has also been shown in
\cite{mcquinn_et_al_2011,altay_et_al_2011,rahmati_et_al_2012}.

Lastly, I studied why this self-shielding occurs at a higher value than one might naively
expect for LLSs. While the hard spectrum from the UVB accounts for most of the
difference, there is also a significant effect from the anisotropic structure of LLSs.
For an absorber with a column density of $\NHI$ in a given direction, I found that on
average, there are lines of sight which have significantly less shielding to the UVB.
This results in the absorber being more ionized than expected from the column density.
Together, these effects result in the onset of self-shielding being pushed to $\NHI =
10^{18}\pcms$. One consequence of this result is that if one can independently constrain
the UVB or the anisotropic structure of LLSs, the other quantity can be constrained by
measuring the column density at which self-shielding kicks in.

I would like to thank the anonymous referee for a thoughtful and thorough report which
improved the quality of the paper. I would like to acknowledge helpful comments from Nick
Gnedin, Andrey Kravtsov, Stephan Meyer, Dan Holz, Tom Witten, Oscar Agertz and Benedikt
Diemer. This work was supported in part by the NSF grant AST-0908063, and by the NASA
grant NNX-09AJ54G. The simulations used in this work have been performed on the Joint
Fermilab - KICP Supercomputing Cluster, supported by grants from Fermilab, Kavli
Institute for Cosmological Physics, and the University of Chicago. This work made
extensive use of the NASA Astrophysics Data System and the arXiv.org preprint server. I
made use of the CAMB code to generate power spectra in the course of this work.

\bibliographystyle{mn2e_long}
\bibliography{citations_LLS_mnras}

\end{document}